# Low frequency and Microwave Magnetoelectric Effects in Thick Film Heterostructures of Lithium Zinc Ferrite and Lead Zirconate Titanate

G. Srinivasan and R. Hayes
Physics Department, Oakland University, Rochester, Michigan 48309-4401, USA
M. I. Bichurin
Department of Physics Engineering, Novgorod State University,
173003 Novgorod, ul. Bolshaya Sankt-Peterburgskaya 41, Russia

## Abstract

Magnetoelectric (ME) coupling at low frequencies and at x-band have been investigated in layered samples containing zinc substituted lithium ferrite and lead zirconate titanate (PZT). Multilayers of $Li_{0.5-x/2}Zn_xFe_{2.5-x/2}O_4$ (LZFO) (x=0-0.4) and PZT were prepared by lamination and sintering of thick films. At low frequencies (10-1000 Hz), the ME voltage coefficient for transverse fields is higher than for longitudinal fields. With Zn substitution in the ferrite, transverse coupling increases to a maximum for x=0.3 and then decreases for higher x. Analysis based on our model for a bilayer implies an efficient magneto-mechanical coupling with Zn substitution which in turn results in strong ME interactions. Microwave ME coupling is studied through measurements of shift in the ferromagnetic resonance field due to an applied electric field. Estimated ME constants from such data are in agreement with our model for a ferrite-PZT bilayer.

PACS Numbers: 75.80.+q; 77.60.+v; 75.70.-i; 77.55.+f

## I. Introduction

Composites are desirable for the synthesis of materials with unique or improved properties. Samples containing piezomagnetic and piezoelectric phases, for example, are product property composites capable of conversion of energies stored in electric and magnetic fields. The field conversion is possible since an applied magnetic field produces a strain in the piezomagnetic phase, which in turn is coupled to the piezoelectric phase, resulting in an induced electric field [1]. The magnetoelectric (ME) coupling is studied by measuring the induced electric field $\delta E$ produced by an applied ac magnetic field $\delta H$. The ME voltage coefficient is given by $\alpha_E = \delta E/\delta H$.

Studies on ME composites were initiated in the early 1970s and were primarily on bulk samples of ferrimagnetic spinel ferrites and piezoelectric barium titanate [1]. Although ferrites are not piezomagnetic, magnetostriction in an ac magnetic field gives rise to pseudo piezomagnetic effects. The bulk composites yielded $\alpha_E$ values that were two to three orders of magnitude smaller than theoretical predictions due to low resistivity for ferrites that creates a leakage current in the sample. Such difficulties are easily overcome in a layered composite since the series electrical connectivity leads to high resistivity and negligible leakage current. For layered samples, theory predicts $\alpha_E$ comparable to bulk composites [2,3]. Ferrite based layered composites studied so far include pure and substituted cobalt ferrite and nickel ferrite – PZT. Samples are usually prepared by cosintering thick films of ferrites and PZT. Studies revealed a maximum $\alpha_E$ on the order of 75-1500 mV/cm Oe, depending on the magnetic and electrical parameters for the two phases [4-8].

Efforts so far on ME composites have mainly focused on low frequency (10 Hz – 1 kHz) phenomena. Studies at microwave frequencies could be performed through the measurement of electric field (E) assisted shift of ferromagnetic resonance lines (FMR) for the ferrite [9]. The shift $\delta H_r$ arises due strain dependence of the resonance field $H_r$ and its magnitude is determined by the piezoelectric and magnetoelastic constants. Thus $\delta H_r$ vs E data could be used to understand the nature of ME couplings and determine ME constants. We investigated the resonance ME effects in bulk composites of 90% YIG-10% PZT and observed the expected shift [10]. But for PZT amounts larger than 20%, the main FMR line broadens and masks any E induced shift. Such difficulties, however, are absent in layered structures since the coupling is essentially an interface phenomenon and FMR line broadening due to nonmagnetic PZT is practically absent.

The focus of this work is an understanding of the effects of magnetic parameters of ferrites on ME coupling in multilayers of $Li_{0.5-x/2}Zn_xFe_{2.5-x/2}O_4$ (LZFO) (x=0-0.4) and PZT. Lithium ferrite, in particular, is appropriate for studies on microwave resonance ME effects because of low losses [11] and thus facilitates accurate determination of $\delta H_r$ and ME constants. We synthesized lithium zinc ferrite-PZT thick film multilayers by sintering films obtained by tape casting. The low frequency ME voltage coefficients were measured for transverse and longitudinal field orientations for frequencies 10 – 1000 Hz. The results provide clear evidence for strengthening of ME effects with the substitution of Zn. We observe a transverse ME coupling that is almost an order of magnitude stronger than the longitudinal coupling due to relative strengths of the piezomagnetic effects. The coefficient $\alpha_E$ increases as the Zn concentration is increased and shows a maximum for x = 0.3. Data analysis based on our model reveals efficient coupling at the ferrite-PZT interface in Zn substituted samples.

For microwave ME effects, ferromagnetic resonance studies at 9.3 GHz were carried out at room temperature on the multilayers. The samples were subjected to a constant electric field perpendicular to its plane and the resonance absorption versus static magnetic field profile was obtained for a series of electric fields. A linear dependence of the field shift on the electric field is evident from the data and has been used for the determination of ME constants of interest. Results on microwave ME coupling have been analyzed using our model for bilayers [9,12].

## 2. Sample Preparation and Characterization

Multilayer structures consisting of alternate layers of ferroelectric and ferromagnetic oxides were prepared from thick films synthesized by the doctor blade technique. First, ferrite powders obtained by standard ceramic techniques and commercial PZT powder (APC 850, American Piezo Ceramics, PA) were used to make a thick suspension in an organic solvent. Films 10-40 $\mu$m in thickness were made from the casts by doctor blade techniques. Details of the film preparation by tape casting are discussed in Ref. [13]. The films were then laminated and sintered at 1200-1500 K. Samples with *n* PZT layers and (*n+1*) ferrite layers (*n*=5-30) were prepared. Structural, magnetic, and dielectric characterizations were done on the multilayers. X-ray diffraction studies indicated the presence of two characteristic sets of reflections, corresponding to the ferrite and PZT, and the absence of new phases. Saturation magnetization measured with a Faraday balance was in agreement with expected values for ferrites. Magnetostriction was measured with the standard strain gage technique. The resistivity, dielectric constant and piezoelectric coupling constants were in agreement with expected values for PZT.

For ME characterization at low frequencies, the samples (5 mm x 5 mm x 0.5 mm) were polished and poled in an electric field. Poling involved heating the sample to 500 K and cooling it back to room temperature in an electric field of 15-30 kV/cm applied perpendicular to the sample plane. Electrodes were deposited on the sample with silver epoxy. For magnetoelectric characterization, the samples were placed in a shielded 3-terminal sample holder and placed between the pole pieces of an electromagnet that was used to apply a static magnetic field H. The required ac magnetic field (10 Hz-1 kHz) $\delta H$ parallel to H was generated with a pair of Helmholtz coils. The resulting ac electric field $\delta E$ perpendicular to the sample plane (direction-*3*) was estimated from the measured voltage (with a lock-in-amplifier). The transverse coefficient $\alpha_{E,31}$ was measured for the magnetic fields (along direction-*1*) parallel to the sample plane and perpendicular to $\delta E$. The longitudinal coefficient $\alpha_{E,33}$ was measured for all the fields perpendicular to the sample plane. (The coefficient $\alpha_E$ is defined for unit thickness of PZT and is related to $\alpha'_E$, the coefficient for unit thickness of the composite by $\alpha'_E = \alpha_E\, t/t'$, where t and t′ are the PZT and composite thickness, respectively.) Magnetoelectric characterization was carried out as a function of frequency of the ac magnetic field, bias magnetic field H and sample temperature.

For microwave ME effects, ferromagnetic resonance studies using a resonance cavity operating at 9.3 GHz were carried out at room temperature on discs of diameter 4 mm and thickness 0.5 mm. The static magnetic field was applied perpendicular to the sample plane and power absorption by the sample was measured as a function of H. The samples were then subjected to a pulsed (2 ms) electric field perpendicular to its plane. The use of pulsed field was necessary to avoid any heating of the sample. The resonance absorption versus static magnetic field profile was obtained for a series of electric fields. The resonance field was thus measured as a function of E and the data was used for estimation of ME constants.

## 3. Results

### (i) Low frequency ME coupling:

Studies were performed on multilayers of LZFO (x=0-0.4) − PZT with $n$=5-30. The ME voltage coefficients were measured as a function of the static magnetic field H, frequency of ac fields, and temperature. Typical H-dependence data are shown in Fig.1 for a LZFO (x=0.2)-PZT containing 16 ferrite and 15 PZT layers with a thickness of 18 µm. Room temperature values of transverse ($\alpha_{E,31}$) and longitudinal ($\alpha_{E,33}$) coefficients measured for 1 Oe ac field at 100 Hz are shown. As the bias field is increased from zero, a rapid increase to a peak value is observed for $\alpha_{E,31}$. With further increase in H, the ME coupling coefficients drop to a minimum. The longitudinal coupling shows a similar H-dependence, but the rise and fall in $\alpha_{E,33}$ with H is less dramatic and the peak value is rather small compared to the transverse case. But the longitudinal coupling is present over a wide H-range. The key observation in Fig.1 is the strong transverse ME coupling. The H dependence in Fig.1 essentially tracks the strength of piezomagnetic coupling $q$=d$\lambda$/dH.

The coupling vanishes when $\lambda$ attains saturation. As discussed later, the features in Fig.1 could be understood in terms of H variation of parallel and perpendicular magnetostriction for the ferrite.

We investigated the effect of Zn substitution on ME coupling. Measurements were carried out on LZFO-PZT multilayers with $n$=15-30 and equal thickness for ferrite and PZT layers. Figure 2 shows the room temperature variation of $\alpha_{E,31}$ with H for x=0-0.3. Data on the longitudinal coupling are not shown since the coupling is quite weak, similar to the data in Fig.1. As x is increased one notices (i) an increase in the slope of $\alpha_E$ vs H at low fields, (ii) the peak in $\alpha_{E,31}$ occurs at progressively increasing H, and (iii) there is substantial increase in the peak value of $\alpha_{E,31}$. In Fig.2 the variation of peak values of $\alpha_E$ with x is shown for transverse fields. As the Zn substitution is increased, one observes a sharp increase in $\alpha_{E,31}$, from 20 mV/cm Oe for x = 0 to 110 mV/cm Oe for x = 0.3. Further increase in x is accompanied by a rapid fall in $\alpha_{E,31}$.

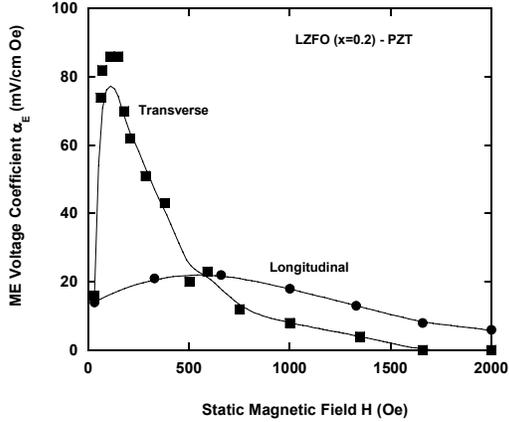

Fig.1: Transverse ($\alpha_{E,31}$) and longitudinal ($\alpha_{E,33}$) magnetoelectric (ME) voltage coefficients as a function of static field H for a multilayer of $Li_{0.5-x/2}Zn_xFe_{2.5-x/2}O_4$ (LZFO) (x=0.2) and PZT. The data at room temperature are for an ac field of 1 Oe at 100 Hz. The transverse coefficient $\alpha_{E,31} = \delta E_3/\delta H_1$ corresponds to H and $\delta$H parallel to each other and to the sample plane (1,2) and the induced electric field $\delta$E measured along the direction-3, perpendicular to sample plane. The longitudinal coefficient $\alpha_{E,33}= \delta E_3/\delta H_3$ is for all the fields along direction-3.

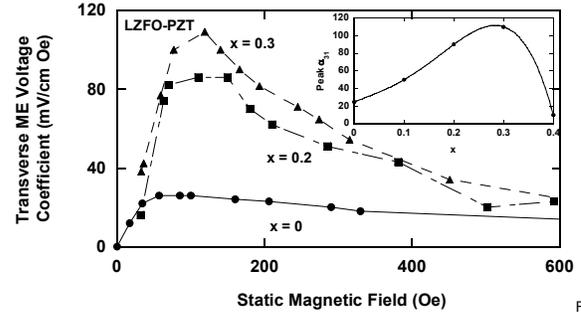

Fig.2: Transverse coefficient $\alpha_{E,31}$ vs H for LZFO (x=0,0.2,0.3) − PZT multilayers. The inset shows the peak $\alpha_{E,31}$ as a function of x.

According to a theoretical model to be discussed in the following section, $\alpha_E$ is expected to be dependent sensitively on the volumes of the magnetostrictive and piezoelectric phases [2,3]. One could control the volume by varying the layer thickness for the two phases. We prepared samples with the volume fraction $v$ for PZT varying from 0.2 to 0.7. We then measured $\alpha_E$ vs H profiles for a series of samples with different $v$ and determined the peak values of $\alpha_E$. Figure 3 shows the variation of peak $\alpha_E$ with $v$ for multilayers with x=0.2. As the PZT volume is increased, one observes a sharp decline in $\alpha_{E,31.}$ The longitudinal coefficient, however, does not show any systematic dependence on *v*. These features are discussed later in terms of our model for a bilayer of ferrite and PZT [2,3]. We also studied the influence of temperature on the strength of ME interactions over the range 100-300 K.

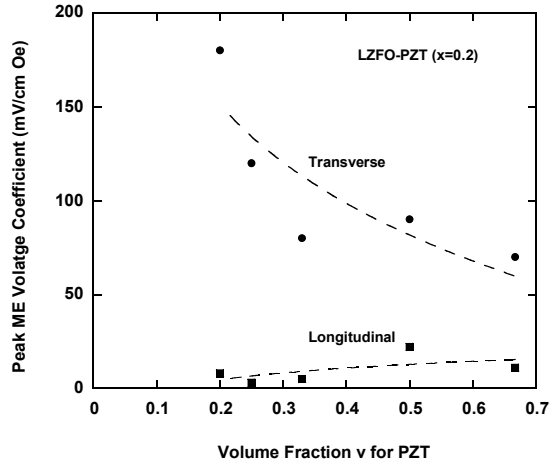

Fig.3: Variation in peak values of transverse and longitudinal ME voltage coefficients with the volume fraction *v* for PZT in a LZFO (x=0.2)-PZT multilayer.

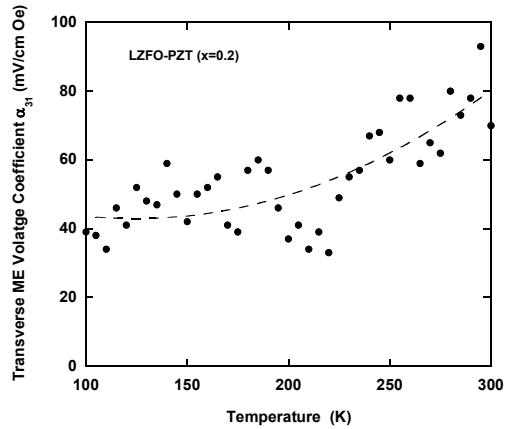

Fig.4: Dependence of peak transverse ME voltage coefficient with temperature for a LZFO (x=0.2)-PZT multilayer.

Figure 4 shows such data on $\alpha_{E,31}$ for a sample with x=0.2 and v=1. A maximum in $\alpha_{E,31}$ is observed at room temperature and it decreases when the temperature is decreased. Detailed temperature dependence of material parameters for both phases is necessary for an understanding of these results.

Now we compare the results in Figs.1 and 2 with ME coupling in similar magnetostrictive-piezoelectric composites. Past studies on bulk composites of cobalt ferrite/nickel ferrite with barium titanate/PZT yielded $\alpha_E$ on the order of 1-130 mV/cm Oe. It is worth noting here that the best values for bulk composites were obtained for samples with modified ferrite in which the resistivity was increased with proper substitutions. Thus the present $\alpha_E$ values for LZFO-PZT compare favorably with best values for bulk composites [1,14]. There has been several recent reports on ME interactions in layered nickel zinc ferrite (NZFO)/PZT and cobalt zinc ferrite (CZFO)/PZT samples [5-8]. The magnitude of $\alpha_E$ in LZFO-PZT is of the same order as in CZFO-PZT, but smaller than for NZFO-PZT. The observations regarding Zn-assisted enhancement on ME coupling (Fig.2) is in agreement with results for layered samples with similar Zn substituted ferrites and PZT. Data analysis based on our bilayer model is provided in Section 4.

*(ii) Microwave ME effects*

Experiments on ME coupling at x-band frequencies differ significantly from the low frequency studies. At low frequencies, we measured the induced electric field produced by the sample in response to an ac magnetic field. For microwave ME coupling, the response of the sample to an *applied electric field* is investigated. Ferromagnetic resonance for the samples was studied at 9.3 GHz studies using a reflection type cavity. The static magnetic field H was applied perpendicular to the sample plane and power absorption vs H profiles were recorded. The samples were then subjected to an electric field E perpendicular to its plane. It was necessary to apply the field in the form of 2 ms pulses in order to eliminate any sample heating. The resonance absorption versus H profile was obtained for a series of electric fields. Figure 5 shows such profiles for LFO-PZT. The results are for a multilayer of LFO (i.e.,x=0)-PZT with 15 micron thick ferrite and PZT layers. The sample contained 16 layers of LFO and 15 layers of PZT.

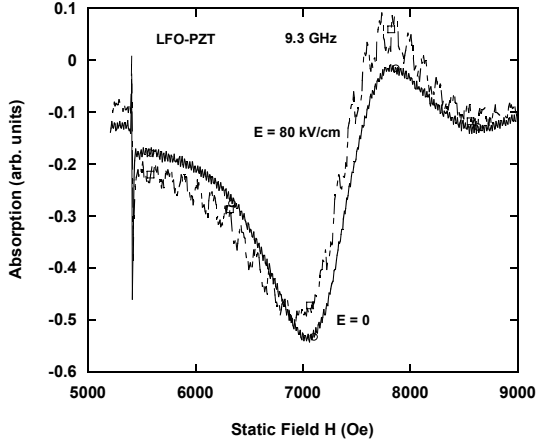

Fig.5

Fig.5: Resonant magnetoelectric effect measured in a multilayer composite of LFO–PZT. The sample contained 16 layers of LFO and 15 layers of PZT. The thickness of each layer is 15 μm. The static field H is perpendicular to the sample plane and 2 ms pulses of E is applied perpendicular to the plane. Absorption vs H profiles are shown for E=0 and E=80 kV/cm.

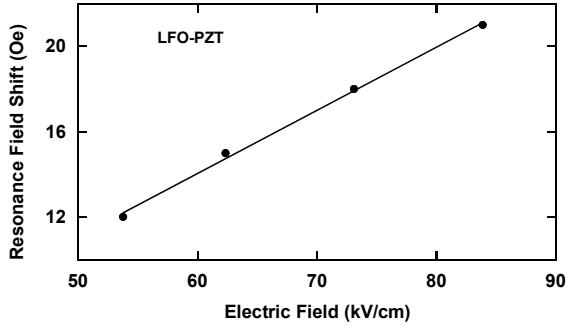

Fig.6: The shift $\delta H_r$ in the ferromagnetic resonance field at 9.3 GHz as a function of E for LFO-PZT.

For E=0, FMR with a line-width ΔH on the order of 300 Oe is observed. With the application of E=80 kV/cm, there is a downshift in the resonance field. Figure 6 shows data on the variation of the shift $\delta H_r$ in the resonance field as a function of E. A linear dependence of the field shift on the electric field is evident from the data and is indicative of the absence of measurable bilinear ME effects. One obtains, from data in Fig. 6, a linear ME coefficient of 0.25 Oe cm/kV. Similar measurements could not be performed for samples with higher x because the FMR absorption was very broad with ΔH ranging from 500 to 1000 Oe. The large ΔH masked any E-induced shift in the resonance field. Analysis of resonance ME effects are discussed in detail in Sec.4.

## 4. Discussion
### (i) Low frequency ME effects

One needs to understand the data in Figs.1 and 2 in terms of variations in the magnetic parameters for the ferrite. Such as approach is justified since the same piezoelectric phase is used in all the composites and the data are exclusively on H-dependence of ME interactions. It is also clear that one needs to focus in on changes in the magnetic parameters of the ferrites when Zn replaces both Li and Fe. We first estimate the magnetoelectric voltage coefficients and their bias magnetic field dependence for comparison with data. Following this, we discuss the possible cause of zinc substitution related enhancement in ME effects.

For theoretical calculation of $\alpha_E$, a basic bilayer of ferrite-PZT is considered. Assuming the structure to be a free body with uniform H and zero electric field in the ferrite, one obtains the following expressions for the transverse and longitudinal ME coefficients [2,3].

Fig.6

$$\alpha_{E,33} = \frac{-2\mu_0 k(1-v)\,{}^p d_{31}\,{}^m q_{31}}{2\,({}^p d_{31})^2 (1-v)k + {}^p\varepsilon^T_{33}\,[({}^p s_{11} + {}^p s_{12})(v-1) - kv({}^m s_{11} + {}^m s_{12})]} \times$$

$$\times \frac{[({}^p s_{11} + {}^p s_{12})(v-1) - kv\,({}^m s_{11} + {}^m s_{12})]}{[\mu_0(v-1) - {}^m\mu_{33}\,v][kv({}^m s_{11} + {}^m s_{12}) - ({}^p s_{11} + {}^p s_{12})(v-1)] + 2\,({}^m q_{31})^2 kv^2} \qquad (1)$$

$$\alpha_{E,31} = \frac{-k(v-1)\ ^{p}d_{31}(^{m}q_{11} + \ ^{m}q_{21})}{(^{m}s_{11} + \ ^{m}s_{12})^{p}\varepsilon^{T}_{33}\ kv + (^{p}s_{11} + \ ^{p}s_{12})\ ^{p}\varepsilon^{T}_{33}\ (1-v) - 2\ (^{p}d_{31})^{2}\ k(1-v)}. \tag{2}$$

where $v = \ ^{p}v/(^{p}v + \ ^{m}v)$ and $^{p}v$ and $^{m}v$ denote the volume of piezoelectric phase and magnetostrictive phase, respectively. Here m denotes the magnetostrictive phase and p the piezoelectric phase, d and q are the piezoelectric and piezomagnetic coupling coefficients, respectively, s is the compliance coefficient and $\varepsilon^{T}$ is permittivity at constant stress. We introduced the coupling parameter k to describe the interface coupling, with k=1 for an ideal interface and k=0 for a frictionless case. (The ME voltage coefficient $\alpha'_{E,33}$ defined for unit length of the composite is related to $\alpha_{E,33}$ through the expression $\alpha_{E,33} = \alpha'_{E,33}/v$.)

The voltage coefficients $\alpha_{E,31}$ and $\alpha_{E,33}$ arise due to piezomagnetic coefficients ($q_{11} + q_{12}$) and $q_{13}$, respectively. Thus one requires the magnitude of $q = \delta\lambda/\delta H$ (where $\lambda$ is the magnetostriction) and its variation with H for the estimation of filed dependence of $\alpha_{E}$. The perpendicular magnetostriction $\lambda_{13}$ and its derivative with H were quite small. Consequently, the longitudinal ME coupling is expected to be weaker than the transverse case, as is the case for data in Fig.1. The discussion to follow is therefore restricted to the transverse ME effect. We determined q-values from data on $\lambda$ vs H for the pure and Zn substituted ferrites. Representative data on the in-plane parallel ($\lambda_{11}$) and perpendicular ($\lambda_{12}$) magnetostriction that are needed for the estimation of $q_{11}$ and $q_{12}$, respectively, are shown in Fig.7 for LZFO bulk samples. The measurements were made with the standard strain gage technique at room temperature on ferrites (1 cm x 1 cm x 0.05 cm) made from thick films. For x = 0 and 0.3, as H is increased we find an increase in the magnitude of $\lambda_{11}$ for fields up to 1.5 kOe where it attains saturation. The in-plane perpendicular magnetostriction $\lambda_{12}$ is small, but positive. There is no noticeable dependence of $\lambda$ on x.

We now use the bilayer model for calculation of $\alpha_{E,31}$ for comparison with the data. The following material parameters were used: $^{p}s_{11} = 15*10^{-12}$ $m^{2}/N$, $^{p}s_{12} = -5*10^{-12}$ $m^{2}/N$, $^{m}s_{11} = 6.5*10^{-12}$ $m^{2}/N$; $^{m}s_{12} = -2.4*10^{-12}$ $m^{2}/N$, $d_{13} = -175$ pm/V, $\mu_{33}/\mu_{0} = 2$, and $\varepsilon_{33}/\varepsilon_{0} = 1750$. The other required parameter, q, was determined from data in Fig.7. Calculated values of $\alpha_{E,31}$ are compared in Fig.7 with the data for LZFO-PZT samples. Results are shown for a series of k-value for samples with x = 0 and 0.3. For ideal interface coupling (k=1) in LFO-PZT, the theory predicts a gradual increase in $\alpha_{E,31}$ with increasing H. A maximum in $\alpha_{E,31}$ is expected for a field of 250 Oe and the ME coefficient drops down to zero for higher H. Upon increasing x from 0 to 0.3, significant theoretical predictions concern a rapid increase in the low field $\alpha_{E,31}$ and a maximum $\alpha_{E,31}$ that is 70% higher than for x = 0. The theoretical $\alpha_{E,31}$ vs H essentially tracks the slope of $\lambda$ vs H in Fig.7.

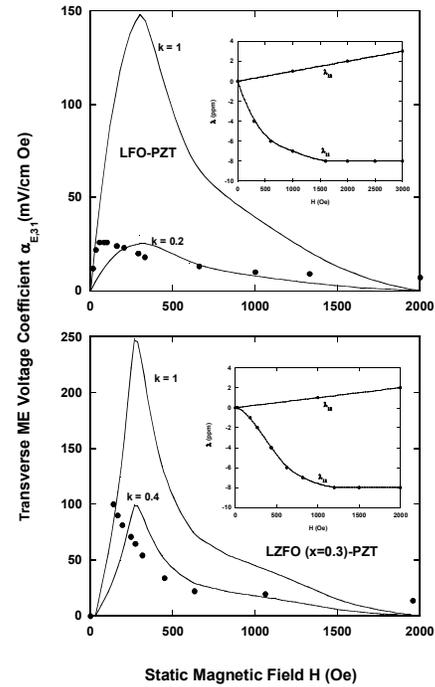

Fig. 7

Fig.7: Comparison of theoretical and measured values of the transverse ME voltage coefficient $\alpha_{E,31}$ for layered samples of LZFO (x=0, 0.2) – PZT. The solid curves are theoretical values for a series for interface coupling parameter k. The solid circles are measured values (Fig.2). The inset shows the H-dependence of the in-plane parallel ($\lambda_{11}$) and perpendicular ($\lambda_{12}$) magnetostriction.

Now we compare the data and theoretical values of $\alpha_{E,31}$ for LZFO-PZT. For x = 0- 0.3, we observe a substantial disagreement between theory for k=1 and data. Neither the magnitude nor the H dependence of calculated $\alpha_{E,31}$ agree with data. Both the predicted maximum $\alpha_{E,31}$ and the corresponding H are higher than measured values. But there is very good agreement between theory and data when the interface coupling k is reduced to 0.2 for x=0 and 0.4 for x=0.3. We, therefore, draw two important inferences from Fig.7: (i) a poor interface coupling in LZFO-PZT and (ii) an improvement in k when Zn is substituted in lithium ferrite. The theory also predicts a dramatic decrease in $\alpha_{E,31}$ with increasing volume of PZT, in agreement with the data in Fig.3.

The results in Fig.7 show several features that are also observed in NZFO-PZT and CZFO-PZT [5-8]. First, for pure ferrites such as CFO − PZT or LFO − PZT, there is total lack of agreement between theory for k=1 and data. A weak interface coupling with k on the order of 0.1-0.2 is evident. Second, the introduction of Zn leads to an enhancement in the strength of ME coupling. There is good agreement between theory for k=0.4-0.6 and data for LZFO − PZT and CZFO − PZT multilayers. Now we discuss the possible cause of poor k-value for pure LFO-PZT. The magnetic parameter of importance here is the Joule magnetostriction that arises due to domain dynamics. Under the influence of a bias field H and ac field $\delta H$, domain wall motion and domain rotation contribute to the Joule magnetostriction. Thus unimpeded domain motion or a high initial permeability is essential for strong magneto-mechanical and ME coupling. High initial permeability ferrites such as nickel or mamganese ferrite is appropriate for ME composites. Thus one can relate poor ME coupling in LFO-PZT and CFO-PZT to the initial permeability $\mu_i$. With the introduction of Zn in the ferrite, however, $\mu_i$ increases [11] and results in enhanced k-values.

*(ii) Microwave ME effects:*

In our theoretical model for high frequency ME effects in bilayer composites, we assumed a bilayer in which the poling axis of the piezoelectric phase coincides with [100] axis of the magnetostrictive phase [9,15]. An external field $E_3$ causes an interface strain due to piezoelectric effect and produces shift of the resonance magnetic field. We obtained the following expression for the field shift

$$\delta H_E = \frac{3\lambda_{100} d_{31} E_3}{M_0[(^p s_{11} + ^p s_{12})(1-\nu) + (^m s_{11} + ^m s_{12})\nu]}$$
$$= A_1 E_3 \quad (3)$$

where $A_1$ is a magnetoelectric constant, $M_0$ is a saturation magnetization and $\lambda_{100}$ is the magnetostriction constant. For lithium ferrite-PZT, using the values $\lambda_{100} = 23\cdot10^{-6}$; $4\pi M_0 =3600\ G$, we obtain the ME constant $A_1$ equals 0.2 Oe/(kV/cm). The estimated value is in excellent agreement than the measured value of 0.25 Oe/(kV/cm) for LFO-PZT multilayers. The ME constant is also a factor of ten higher than for bulk YIG-PZT composites [10], primarily due to high magnetostriction for LFO.

## 5. Conclusions

The nature of low frequency and microwave ME coupling and its dependence on composition has been investigated in layered samples of lithium zinc ferrite and PZT. At low frequencies, the transverse coupling is the dominant ME interaction and is almost an order of magnitude stronger than longitudinal ME voltage coefficient. Zinc substitution in lithium ferrite results in improved magneto-mechanical and ME couplings. Information on microwave ME effects has been obtained through the effects of external electric field on ferromagnetic resonance for the ferrite. The ME constant estimated from such data is indicative of strong high frequency ME interactions, in agreement with theory.

## Acknowledgments

The research is supported by a grant from the National Science Foundation (DMR-0302254).